\documentstyle[12pt]{article}
\begin{document}
\title{Lattice Bosons}
\vspace{ 1 cm }
\begin{center}
\author{$J.Chakrabarti^{a,1},A.Basu^{a} and B.Bagchi^{b,2}$}
\date{}
\maketitle
\end{center}
a. Department of Theoretical Physics \\
   Indian Association for the Cultivation of Science \\
   Calcutta-700032   INDIA \\
b. Department of Applied Mathematics \\
   University of Calcutta \\
   92 Acharya Prafulla Chandra Road \\
   Calcutta-700009   INDIA\\

\begin{abstract}

Fermions on the lattice have bosonic excitations generated from the underlying 
periodic background. These , the lattice bosons , arise near the empty band or when

the bands are nearly full. They do not depend on the nature of the interactions
and exist for any fermion-fermion coupling .We discuss these lattice boson solutions for the Dirac Hamiltonian.

\end{abstract}

\vspace{0.5cm}



\vspace{2cm}
1.Electronic address: tpjc@mahendra.iacs.res.in \\
2.Electronic address: bbagchi@cucc.ernet.in\\

\newpage
\section{Introduction}

The fermions often have boson solutions. These bosons, made usually from the
fermion degrees ,exist in many systems [1]. 
We present in this work the boson solutions that arise from the
background lattice.They do not arise from the interactions, but are related to the spatial
geometrical features, such as the periodicities, of the underlying lattice  on
which the fermions reside.Because of this geometrical origin these bosons, we
believe, are robust and exist for any value of coupling.These,the lattice bosons,are the subject of this work.
\par
It is worth recalling the many-body fermion states in low dimensions are 
eigenstates of boson operators.These ideas of bosonisation have received wide
support over the years[2-6].
\par
Here, in this work , we illustrate our ideas on the one-dimensional equispaced
lattice. We find the fermion algebra that is represented by this lattice.This 
algebra of fermions shows the curious feature that for low values of fermion
filling on the lattice some of the generators behave as bosons. Interestingly,
the same is true in the "insulating" region where the fermion states are nearly
full.The boson interpretations in these two limits,the nearly empty lattice(nel),
 and the nearly filled lattice(nfl),are different.The roles of the creation
and the destruction operators are reversed.These interpretations do not depend
on the underlying Hamiltonian or the nature of the fermion interactions.
\par

The lattice bosons in the nel are made of coherently superposed fermion pairs.In the "insulating " region, i.e. in the nfl,coherently superposed hole pairs create the lattice bosons.The usual Hamiltonian of the fermions is recast in terms of these bosons in the nel and the nfl regions.The new Hamiltonian in the lattice boson variables is diagonalised to get the boson spectrum.
\par
To begin we introduce on the one-dimensional equispaced lattice , with periodic
boundary conditions,the set of generators,made of fermionic objects, that close
under commutations.This algebraic structure, represented by the background
lattice, arises solely from the anticommuting properties of the fermions.
It does not depend on any Hamiltonian or on the interactions.We illustrate
the algebra on the simple 6-point lattice.
Following from this algebraic structure we identify the generators that are
bosonic.We show that for low  values of fermion filling a set of generators
satisfy bosonic commutation rules.As the filling increases the boson approximation gets
worse.Interestingly ,near maximum filling , i.e. near the "insulating" domain
the boson modes return,but now with the creation and the destruction operators
reversing their roles. This leads us to interpret the bosons as coherent
superposition of the hole pairs.
\newline
With this structure in place we take up a simple Hamiltonian and look for
its lattice boson solutions. Our choice is the Dirac Hamiltonian for fermions of mass.

We set up the lattice boson algebra and solve for the boson eigenstates. These 
are good solutions in the nel and the nfl regions.    

\section{The Lattice Boson Algebra}

Consider  the lattice of 2N equispaced points with periodic boundary conditions.
Let $C_{n}^{\dagger}$ and $C_{n}$ denote the creation and the annihilation operators
of the fermion on site denoted by the index n.Clearly,
\vspace{0.2cm}
\begin{equation}
\{C_{n},C_{m}^{\dagger}\}=\delta_{nm}
\end{equation}

\begin{equation}
\{C_{n}^{\dagger},C_{m}^{\dagger}\}=\{C_{n},C_{m}\}=0
\end{equation}

\vspace{0.2 cm } 
Consider the generators,
\vspace{0.2cm}
\begin{equation}
e_{+l}=\sum C_{n}^{\dagger}C_{n+l}^{\dagger},
\end{equation}
\vspace{0.2 cm}
where l takes values 1,2,3...   For the lattice of 6 points the generators  
read :
\vspace{0.2cm}
\begin{eqnarray}
e_{+1}=c_{1}^{\dagger}c_{2}^{\dagger}+c_{2}^{\dagger}c_{3}^{\dagger}+
c_{3}^{\dagger}c_{4}{\dagger}+c_{4}^{\dagger}c_{5}^{\dagger}+c_{5}^{\dagger}
c_{6}^{\dagger}+c_{6}^{\dagger}c_{1}^{\dagger}\\
e_{+2}=c_{1}^{\dagger}c_{3}^{\dagger}+c_{2}^{\dagger}c_{4}^{\dagger}
+c_{3}^{\dagger}c_{5}^{\dagger}+c_{4}^{\dagger}c_{6}^{\dagger}
+c_{5}^{\dagger}c_{1}^{\dagger}+c_{6}^{\dagger}c_{2}^{\dagger}\\
e_{+3}=c_{1}^{\dagger}c_{4}^{\dagger}+c_{2}^{\dagger}c_{5}^{\dagger}+c_{3}
^{\dagger}c_{6}^{\dagger}+c_{4}^{\dagger}c_{1}^{\dagger}+c_{5}^{\dagger}
c_{2}^{\dagger}+c_{6}^{\dagger}c_{3}^{\dagger}
\end{eqnarray}
\vspace{0.2cm}
and so on.
\newpage
 But ,interestingly, $e_{+3}=0$ , since the last three terms of $e_{+3}$ ,using (2),
are negatives of the first three . Further $e_{+4}$ and $e_{+5}$ 
can also be shown to be negatives of $e_{+2}$ and $e_{+1}$ respectively . So 
for the lattice of 6(i.e. (2N)) points the number of independent generators($e_{+l}$)
are 2 (i.e. (N-1)), corresponding to l=1 and l=2.
Consider now the conjugates of $e_{+l}$ . Denoted by $e_{-l}$ ,for this case of
the 6-point-lattice,   these are:
\vspace{0.2cm}
\begin{eqnarray}
e_{-1}=-[c_{1}c_{2}+c_{2}c_{3}+c_{3}c_{4}+c_{4}c_{5}+c_{5}c_{6}+c_{6}c_{1}]\\
e_{-2}=-[c_{1}c_{3}+c_{2}c_{4}+c_{3}c_{5}+c_{4}c_{6}+c_{5}c_{1}+c_{6}c_{2}]
\end{eqnarray}
\vspace{0.2cm}
Now if we calculate the commutator of $e_{+1}$ and $e_{-1}$ ,we get:
\vspace{0.2cm}
\begin{equation}
[e_{-1},e_{+1}]=6 - 2\sum c_{n}^{\dagger}c_{n}+ \sum (c_{n}^{\dagger}c_{n+2}
+h.c.)
\end{equation}
\vspace{0.2cm}
where we have used the anticommutators ( 1 ).The quantities on the r.h.s. of (9)
\vspace{0.2cm} 
\begin{equation}
h_{0}= \sum c_{n}^{\dagger}c_{n}
\end{equation}
\begin{equation}
h_{2}= \sum (c_{n}^{\dagger}c_{n+2}+h.c.)
\end{equation}
\vspace{0.2cm}
are the fermion number operators, $h_{0}$ ,and the hopping operator, $h_{2}$.
\newpage
The quantity 6(i.e.,2N) is just the total number of points on the lattice. Similarly,
\vspace{0.2cm}
\begin{equation}
[e_{-2},e_{+1}]=-h_{1}+h_{3}
\end{equation}
\vspace{0.2cm}
where the $h_{i}$ is defined as $\sum (c_{n}^{\dagger}c_{n+i}+h.c.)$; the $h_{1}$ and  the $h_{3}$ are
two hopping operators.It is then easy to check that :
\vspace{0.2cm}
\begin{equation}
[e_{\pm l},h_{0}]=\mp 2e_{\pm l}
\end{equation}
\begin{equation}
[e_{\pm 1},h_{1}]=\mp e_{\pm 2}; [e_{\pm 1},h_{2}]=\pm e_{\pm 1};
[e_{\pm 1},h_{3}]=\pm e_{\pm 2}
\end{equation}
\begin{equation}
[e_{\pm 2},h_{1}]=\mp e_{\pm 1};[e_{\pm 2},h_{2}]=\pm e_{\pm 2};
[e_{\pm 2},h_{3}]=\pm e_{\pm 1}
\end{equation}
\vspace{0.2cm}
Further,
\vspace{0.2cm}
\begin{equation}
[e_{+i},e_{+j}]=[e_{-i},e_{-j}]=[h_{i},h_{j}]=0
\end{equation}
\vspace{0.2cm}
for all i and j. 
Thus $e_{\pm 1},e_{\pm 2},h_{0},h_{1},h_{2},h_{3}$,form a closed algebra for
the 6-point lattice.
The generalisation to the case of the 2N lattice is straightforward.For 
this general case the number of independent $e_{+l}$ generators are (N-1).
The number of independent $h_{i}$ generators are (N+1).
\newpage
The structure of
the algebra is:
\vspace{0.2cm}
\begin{equation}
[e_{+l},e_{-l}]=2N-2h_{0}+h_{2l}
\end{equation}
\begin{equation}
[e_{+l},e_{-l^{\prime}}]=\sum \alpha_{ll^{\prime}}^{j}h_{j}\:\:\:\:\:\:\:\:\:\:
 (l\neq l^{\prime})
\end{equation}
\begin{equation}
[e_{\pm l},h_{0}]= \mp 2e_{\pm l}
\end{equation}
\begin{equation}
[e_{\pm l},h_{i}]=\sum \beta_{li}^{j}e_{\pm j}
\end{equation}
\begin{equation}
[e_{+l},e_{+l^{\prime}}]=[e_{-l},e_{-l^{\prime}}]=[h_{i},h_{j}]=0
\end{equation}
\vspace{0.2cm}

Note that the sums over j in (18) and (20) run over only a small number of j values. That is $\alpha_{ll^{\prime}}^{j}$ and $\beta_{li}^{j}$ are non-zero only for one or two values of j for given $ll^{\prime}$ and li.The fermion number h$_0$ does not appear on the rhs of (18).

\section{The Boson Interpretation}
Consider now the commutator (17) of
\vspace{0.2cm} 
\begin{equation}
[e_{-l},e_{+l}]=2N-2h_{0}+h_{2l}
\end{equation}
\vspace{0.2cm}
where $h_{0}$ is the fermion number operator. 
The first term, 2N , is the size
of the lattice. For small values of fermion filling i.e. when the band is nearly 
empty $h_{0}$ is way small compared to 2N.
\par
Further the operator $h_{2l}$
given by :
\vspace{0.2cm}
\begin{equation}
h_{2l}=\sum( C_{n}^{\dagger}C_{n+2l}+ h. c.)
\end{equation}
\vspace{0.2cm}
in its diagonal form reads:
\vspace{0.2cm}
\begin{equation}
h_{2l}=\sum 2 cos 2kla C_{k}^{\dagger}C_{k}
\end {equation}
\vspace{0.2cm}
Thus the  value of $h_{2l}$ for the state of the single fermion is
 bounded by $\pm 2 $.For the small number of fermions in the lattice(nel)
the value of $h_{2l}$ is again way small compared to the first term 2N.
Therefore equation (22) in the nel region becomes
\vspace{0.2cm} 
\begin{equation}
[e_{-l},e_{+l}]=2N
\end{equation}
\vspace{0.2cm}
For the normalised generators $\frac{1}{\sqrt 2N} e_{\pm l}$  the r.h.s. of (25)
becomes 1.Consider now the commutators $[e_{-l},e_{+l'}]$ with different l and $l^{\prime}$ .From (18) we get 
$[e_{-l},e_{+l'}]=\sum \alpha_{ll^{\prime}}^{j}h_{j}$ (in the sum 
$j \neq 0$).  
The generators $h_{i}$ are all simultaneously diagonalisable.All of them have 
eigenvalues bounded by 2 for the single fermion .(See also the discussions following the eqn (21)).For the normalized generators
we ,therefore, get:
\vspace{0.2cm}
\begin{equation}
[e_{-l},e_{+l'}]=0\hspace{2mm}  when\hspace{2mm}  l\neq l'
\end{equation}
\vspace{0.2cm}
\newpage
Taken together (25) and (26) gives for all values of l and 
$l^{\prime}$ the relation:
\vspace{0.2cm}
\begin{equation}
[e_{-l},e_{+l'}]=\delta _{l,l'}
\end{equation}
\vspace{0.2cm}
These are therefore the creation and the destruction operators of bosons.
Consider now  the insulating limit , when the lattice is nearly full(nfl).
In the eqn (17) the value of $h_{0}$ is about 2N, so that combining the first two terms
on the r.h.s. of (17) we get
\vspace{0.2cm} 
\begin{equation}
[e_{-l},e_{+l}]=-2N
\end{equation}
\vspace{0.2cm}
as for the nfl the value of  the $h_{i}$ are finite ,near zero.
Thus if the creation and the annihilation operators are interchanged ,the 
same bosonic commutator relation reappears.
In the insulating limit ,therefore, the coherent superposed pairs of holes 
created by $e_{-l}$ on the insulator is the lattice boson creation operator.

\newpage    
\section{The Dirac Hamiltonian}   
Consider the Dirac Hamiltonian with mass on the one-dimensional equispaced
lattice of 2N sites. The continuum Hamiltonian is[7,8] :
\vspace{0.2cm}
\begin{equation}
H=- i\psi^{\dagger}(\alpha.\partial)\psi
\end{equation}
\vspace{0.2cm}
with $\psi$ being the two-component wave-function 
$( ^{\psi_{1}}_{\psi_{2}})$  and the Dirac matrices
of interest are $\gamma_{0}=\sigma_{3}$ ; $\alpha=\gamma_{5}=\sigma_{1}$
given by:
\begin{equation}
\left(\begin{array}{lr}1&0\\0&-1
\end{array}
\right)and
\left(\begin{array}{lr}0&1\\1&0
\end{array}
\right)
\end{equation}
respectively.
If we choose 
\begin{displaymath}
 \psi_{\pm}=\frac{1}{2}(1\pm \gamma_{5})\psi
\end{displaymath}
and then write $c=\psi_{1}+\psi_{2}$, $b=\psi_{1}-\psi_{2}$ and define the derivative as:
\begin{displaymath}
\partial c=\frac{c_{n+1}-c_{n-1}}{2a}
\end{displaymath}
,a being the lattice spacing,the Dirac Hamiltonian on the lattice takes the
form:
\vspace{0.2cm}
\begin{equation}
H=H_{c}+H_{b}+H_{m}=i\sum (c_{n}^{\dagger}c_{n+1}-h.c.)-i\sum (b_{n}^{\dagger}
b_{n+1}-h.c.)+m\sum (c_{n}^{\dagger}b_{n}+h.c.)
\end{equation}
where we have included the mass term as  well.
\vspace{0.2cm}

In arriving at (31) we have set the lattice spacing to 1/2.
\newpage
 The choice of a is not
important in our considerations.Our results are good for any value of a.
The lattice boson generators for this Hamiltonian are now written as :
\vspace{0.2cm}
\begin{equation}
e_{+l}^{c}=\sum c_{n}^{\dagger}c_{n+l}^{\dagger};\hspace{2mm}
e_{-l}^{c}=-\sum c_{n}c_{n+l}
\end{equation}
\vspace{0.2cm}
and
\vspace{0.2cm}
\begin{equation}
e_{+l}^{b}=\sum b_{n}^{\dagger}b_{n+l}^{\dagger};\hspace{2mm}
e_{-l}^{b}=-\sum b_{n}b_{n+l}
\end{equation}
\vspace{0.2cm}
The other generators $h_{i}^{c}$ and $h_{i}^{b}$ may be constructed in analogy
with h$_i$ (defined immediately following eqn (12)).
In the discussions that follow it is more convenient to consider the linear
combinations
\vspace{0.2cm}
\begin{equation}
E_{\pm l}^{\pm }=\frac{1}{\sqrt{2}}(e_{\pm l}^{c}\pm e_{\pm l}^{b})
\end{equation}
\vspace{0.2cm}
For the case of the Dirac Hamiltonian (31) the lattice boson generators (34)
are not complete, but have to  include the following further ones:
\vspace{0.2cm}
\begin{equation}
d_{+l}^{1}=\sum c_{n}^{\dagger}b_{n+l}^{\dagger}\hspace{2mm};
d_{+l}^{2}=\sum b_{n}^{\dagger}c_{n+l}^{\dagger}
\end{equation}
\vspace{0.2cm}
along with their conjugates , $d_{-l}^{1}$ and $d_{-l}^{2}$. Once again 
the linear combinations
\vspace{0.2cm} 
\begin{equation}
D_{\pm l}^{\pm }=\frac{1}{\sqrt{2}} (d_{\pm l}^{1} \pm d_{\pm l}^{2})
\end{equation}
\vspace{0.2cm} 
that turn out to be of interest.
\newpage
 Working through the algebra we find that
\vspace{0.2cm}
\begin{equation}
[E_{-l}^{j},E_{+l^{\prime}}^{j^{\prime}}] =  2N \delta _{ll^{\prime}}
\delta_{jj^{\prime}}
\end{equation}
\vspace{0.2cm}
\begin{equation}
[D_{-l}^{j},D_{+l^{\prime}}^{j^{\prime}}]  =  2N \delta_{ll^{\prime}}
\delta_{jj^{\prime}}
\end{equation}
\vspace{0.2cm}
\begin{equation}
[E_{\pm l}^{j},D_{\pm l^{\prime}}^{j^{\prime}}]  =  0
\end{equation}
\vspace{0.2cm}
where j,j' take values + and - .
In the nel region we are led ,therefore , to interpret the $E_{\pm l}^{j}$
and $D_{\pm l}^{j}$ as the lattice boson generators.Note the eqn (39) follows  the same approximation as the eqn(26).The operators must be normalised as in (26).
The analogy follows equally well for the nfl region ,where the right-hand 
sides of (37) and (38) reverse the signs.Thus the superposed hole pairs created on the
insulating region by the actions of $E_{-l}^{j}$ and $D_{-l}^{j}$ are the 
lattice bosons.
To solve  for  these lattice boson modes for the Dirac Hamiltonian let us 
calculate the commutators of $E_{\pm l}^{j}$ and $D_{\pm l}^{j}$ with the
Hamiltonian (31).These give :
\vspace{0.2cm}
\begin{equation}
[H_{c}+H_{b},E_{l}^{\pm}]=0
\end{equation}
\vspace{0.2cm}
\begin{equation}
[H_{m},E_{l}^{+}]=2mD_{l}^{+}
\end{equation}
\vspace{0.2cm}
\begin{equation}
[H_{m},E_{l}^{-}]=0
\end{equation}
\vspace{0.2cm}
while
\vspace{0.2cm}
\begin{equation}
[H_{c}+H_{b},D_{l}^{+}]=2i[D_{l+1}^{-}-D_{l-1}^{-}]
\end{equation}
\vspace{0.2cm}
\begin{equation}
[H_{c}+H_{b},D_{l}^{-}]=2i[D_{l+1}^{+}-D_{l-1}^{+}]
\end{equation}
\vspace{0.2cm}
\begin{equation}
[H_{m},D_{l}^{+}]=2mE_{l}^{+}
\end{equation}
\vspace{0.2cm}
\begin{equation}
[H_{m},D_{l}^{-}]=0
\end{equation}
\vspace{0.2cm}
The equivalent lattice boson Hamiltonian thus is:
\vspace{0.2cm}
\begin{equation}
H_{B}=2i[\sum D_{+(l+1)}^{-}D_{-l}^{+}+\sum D_{+(l+1)}^{+}D_{-l}^{-}]+2m\sum
D_{+l}^{+}E_{-l}^{+}+ h.c.
\end{equation}
\vspace{0.2cm}
The Hamiltonian $H_{B}$ is equivalent to (31) in the sense that they produce
the same commutators for the lattice boson generators when we use the boson
commutators (37-39).To diagonalize $H_{B}$ we first carry out the transform:

\vspace{0.2cm}
\begin{equation}
E_{\pm l}^{\pm}=\frac{1}{\sqrt{L}}\sum E_{\pm}^{\pm}(q)e^{\mp iql}
\end{equation}
and

\begin{equation}
D_{\pm l}^{\pm}=\frac{1}{\sqrt{L}}\sum D_{\pm}^{\pm}(q)e^{\mp iql}
\end{equation}
\vspace{0.2cm}
where L is the number of independent points on the l-lattice, i.e.L=(N-1).
[see the discussion below eqn. (11)].
The transformed $H_{B}$ reads :
\vspace{0.2cm}
\begin{equation}
H_{B}= \sum {-}4 sinq [D_{+}^{-}(q)D_{-}^{+}(q)+D_{+}^{+}(q)D_{-}^{-}(q)]
+2m\sum (D_{+}^{+}(q)E_{-}^{+}(q)+E_{+}^{+}(q)D_{-}^{+}(q))
\end{equation}
\vspace{0.2cm}
The Hamiltonian matrix in the space of $E_{\pm}^{+}(q)$ and $D_{\pm}^{\pm}(q)$  reads as:
\vspace{0.2cm}
\begin{equation} 
\left( \begin{array}{lcr}0 & a & 2m \\ a & 0 & 0 \\ 2m & 0 & 0 \end{array}
\right)
\end{equation}
where a=-4sinq .
\vspace{0.2cm}
\par 
The above Hamiltonian matrix (51) is diagonalized to give us the
eigenstates of the lattice bosons with eigenvalues 0 and 
$\pm \sqrt{a^{2}+2m^{2}}$ in the nel region.In the nfl region, because
of the minus sign in(28), there is an overall minus sign for $H_{B}$
of (47).Thus the eigenvalues in the nfl region just reverse their
signs.
\par 
From eqn.(40) note that $E_{\pm l}^{-}$
are generators of symmetries of the massless Dirac Hamiltonian.
Prior to the insertion of mass term, $E_{\pm l}^{\pm}$
are symmetries.The mass term keeps the $E_{\pm l}^{-}$
symmetries but breaks $E_{\pm l}^{+}$.The $E_{\pm l}
^{+}$ mixes with $D_{\pm l}^{+}$.The $E_{\pm l}^{-}$
generate zero mass  bosonic excitations that normalises
the ground state. Since the $E_{\pm l}^{-}$ modes remain decoupled they have not explicitly appeared
in our equation(47).
It is important to point out that the number of independent $D_{l}^{\pm}$ generators differ a bit from
the number of $E_{l}^{\pm}$.
\par
We have shown[see below eqn.(16)] that the number of independent 
$e_{+l}$ generators for the lattice of size 2N is
given by N-1.For the case of the Dirac Hamiltonian 
of mass we have $e_{+l}^{c}$ and $e_{+l}^{b}$
making a total of 2(N-1) independent generators. There
are , from (34) , 2(N-1) generators of type $E_{+l}^{\pm}$.
\par
The cases of $d_{+l}^{1,2}$ are roughly the same,except
that $d_{+0}^{1,2}$ are independent and non-zero.They
add extra degrees. For the $D_{+l}^{\pm}$, made of the
$d_{+l}^{1,2}$,eqn. (36), the $D_{+0}^{+}=0$.
This is because 
\vspace{0.2cm}
\begin{equation}
D_{+0}^{+}=d_{+0}^{1}+d_{+0}^{2}=\sum c_{n}^{\dagger}
b_{n}^{\dagger}+b_{n}^{\dagger}c_{n}^{\dagger}=0
\end{equation}
\vspace{0.2cm}
from the anticommutators $\{c_{n}^{\dagger},b_{m}^{\dagger}\}=0$ Thus the number
of $D_{l}^{+}$ and $E_{l}^{+}$ are identical.
\par 
The case of $D_{+0}^{-}$ is as follows :
\begin{equation}
D_{+0}^{-}=d_{+0}^{1}-d_{+0}^{2}=2\sum c_{n}^{\dagger}
b_{n}^{\dagger}
\end{equation}
This non-zero $D_{+0}^{-}$ appears in (47 ).Thus in the
hopping part of the effective boson Hamiltonian,$H_{B}$,
these $D_{\pm 0}^{-}$ appear at the end point of the 
l-lattice for the D-operator.This end point is neglected in our diagonalisation of $H_{B}$.Adding this
end point ,for reasonably large values of N, does not alter our results.
  
\section{Discussions}
The fermions anticommute. This leads to fermion composites that behave as
bosons on the lattice.The lattice background gives rise to the set of bosons
that we call the lattice bosons. They exist near the empty lattice (nel) when 
the number of fermions are small.In the insulating region (nfl) these                    
lattice bosons reappear as coherently superposed hole pairs on the insulator[9].
\par
Since these bosons depend on the fermion algebra, they exist quite independent
of the fermion-fermion interaction, or on the interactions of the fermions with
the lattice background. In particular they exist even for small values of 
couplings.To the lowest order ,we have seen from the model of the Dirac fermions
the lattice boson wavefunctions overlap,leading to the hopping over the 
l-lattice.This Hamiltonian is readily diagonalisable.   
It is to be recognised that the mass term in (50) could come from the 
interactions.It could arise from the fermions coupling to the Higgs boson
or from parts of the QCD that lead to the dynamical mass generation and the
chiral symmetry breaking [10].
\par
To  conclude ,we have  shown  that  for  fermion  systems  there  are  boson
excitations  that  arise  from  the  background lattice.   


\end{document}